\documentstyle[preprint,aps,epsfig]{revtex}
\draft
\begin{document}
\bibliographystyle{prsty}
 
\preprint{RUB-TPII-11/95}
\title{Update of the stranger story: The strange vector
form factors of the nucleon in the SU(3) chiral quark-soliton model}

\author{Hyun-Chul Kim
\footnote{E-mail address: kim@hadron.tp2.ruhr-uni-bochum.de},
Teruaki Watabe 
\footnote{E-mail address: watabe@hadron.tp2.ruhr-uni-bochum.de},
and Klaus Goeke
\footnote{E-mail address: goeke@hadron.tp2.ruhr-uni-bochum.de}}

\address{
Institut f\"ur Theoretische  Physik  II, \\  Postfach 102148,
Ruhr-Universit\" at Bochum, \\
 D--44780 Bochum, Germany  \\
       }
\date{March 1996}
\maketitle
\begin{abstract}
The strange vector form factors are evaluated 
for $Q^2=0$ and $Q^2=1\ \mbox{GeV}^2$ in the framework
of the SU(3) chiral quark-soliton model (or semi-bosonized SU(3) 
Nambu-Jona-Lasinio model).   The rotational $1/N_c$ 
and $m_s$ corrections are taken into account up to linear order.  
The mean-square strange radius $\langle r^{2}\rangle^{Sachs}_{s}=-0.35\;
\mbox{fm}^2$ and the strange magnetic moment $\mu_s = -0.44\;\mu_N$
are obtained.  The results are compared with several different models.  
\end{abstract}
\pacs{PACS: 12.40.-y, 14.20.Dh}

\section{Introduction}
The strangeness content of the nucleon has been under a great deal 
of discussions for well over a decade.
A few years ago, the European Muon Collaboration (EMC)~\cite{ashman} 
measured the spin structure function of the
proton in deep inelastic muon scattering and
showed that there is an indication of
a sizable strange quark contribution. 
This remarkable result has been confirmed by following experiments
of the Spin Muon Collaboration (SMC)~\cite{SMC1,SMC2},
E142 and E143 collaborations~\cite{E142,E143}.

Another experiment conducted at Brookhaven~\cite{ahrens} 
(BNL experiment 734)
measuring the low-energy elastic neutrino-proton scattering
came to the more or less same conclusion.  
 Kaplan and Manohar~\cite{km} 
showed how elastic $\nu p$ and $ep$ scatterings can be used 
to extract not only the $G_{1}$ form factors of the $U(1)_A$ current
but also the $F_2$ form factors of the baryon number current
and furthermore how the strange quark matrix elements
$\langle p | \bar{s} \gamma_\mu \gamma_5s|p \rangle $ and 
$\langle p | \bar{s} \gamma_\mu s | p \rangle $ can be evaluated.
 Following these suggestions, 
Garvey {\em et al.}~\cite{garvey1} reanalyzed the above-mentioned
$\nu p$ elastic scattering experiment and determined proton
strange form factors in particular at $Q^2=0$, pointing
out the shortcomings of the analysis done by Ref.~\cite{ahrens} .
The best fit of Ref.~\cite{garvey1} with the smallest $\chi^2$ tells
$F^{s}_{1}=0.53\pm 0.70$ and $F^{s}_{2}=-0.40\pm 0.72$.
By comparing the different $Q^2$ dependence of 
$d\sigma/dQ^2(\nu p)$ to $d\sigma/dQ^2(\bar{\nu} p)$,
Garvey {\em et al}. favor $F^{s}_{1} (Q^2) > 0$ and
$F^{s}_{2} (Q^2) < 0$.              
However, these form factors are 
experimentally unknown to date and have no stringent 
and concrete constraints on their $Q^2$--dependence yet.  
There are various proposals and experiments in progress
(see Refs.~\cite{musolfetal,bm} for details).
All these considerations lead to the conclusion that,
in contrast to the naive quark model, it is of great importance
to consider strange quarks in the nucleon seriously.

There have been several theoretical efforts to describe the strange 
form factors of the nucleon.  The first attempt
was performed by Jaffe~\cite{jaffe}.  Jaffe took advantage of
Ref.~\cite{hoehler}, {\em i.e.}
the pole fit analysis based on dispersion theory and
estimated the mean-square strange radius and magnetic moment
of the nucleon:$\langle r^2\rangle^{Dirac}_{s} 
= 0.16\pm 0.06 \ \mbox{fm}^2$, $\mu_s = -0.31\pm 0.09 \ \mu_N$.
 More recently, 
Hammer {\em et al.}~\cite{Hammeretal} updated Jaffe's pole-fit 
analysis of the strange vector form factors,
relying upon a new dispersion theoretic analysis of the nucleon 
electromagnetic form factors.  In fact, Hammer {\em et al.} improved
Jaffe's prediction, giving $\mu_s = -0.24\pm 0.03 \ 
\mu_N$ and $\langle r^2\rangle^{Dirac}_s = 0.21 \pm 0.03\ \mbox{fm}^2 $.
A noticeable point of the pole-fit analysis is that it has the different
sign of the strange electric radius, compared with almost other models.

Another interesting approach is the kaon-loop calculation.
The main idea of the kaon-loop calculation is that the strangeness
content of the nucleon exists as a pair of $K\Lambda$ or $K\Sigma$
components.
 Koepf {\em et al.}~\cite{Koepfetal} first evaluated $\mu_s$ and 
$\langle r^2\rangle^{Dirac}_s$, considering the possible kaon loops
relevant for the strange vector form factors.  
 However, Ref.~\cite{Koepfetal}
failed to include {\em seagull} terms which are 
essential to satisfy the Ward-Takahashi identity in the vector 
current sector.  Musolf {\em et al.}~\cite{musolfa} 
added these seagull terms and obtained
$\mu_s = -(0.31\rightarrow 0.40) \ \mu_N$ and 
$\langle r^2\rangle^{Dirac}_s = -(6.68\rightarrow 6.90)\times10^{-3}
\ \mbox{fm}^2 $. 
The prediction of $\langle r^2\rangle^{Dirac}_s$ in the kaon-loop 
calculation is found to be much smaller than the pole-fit analysis.  
To reconcile the conflict between the pole-fit analysis and the 
kaon-loop calculation, Refs.~\cite{Cohenetal,forkel} 
suggested the combination of the vector meson dominance (VMD) and
$\omega-\phi$ mixing in the vector-isovector channel 
with the kaon-loop calculation.
The value of $\langle r^2\rangle^{Dirac}_s$ in Ref.~\cite{Cohenetal} 
appeared to be larger than that of the kaon-loop calculation but
still conspicuosly smaller than that of the pole-fit analysis:
$\langle r^2\rangle^{Dirac}_s=-(2.42\rightarrow 2.45)\times 10^{-2}
\ \mbox{fm}^2 $.  Ref.~\cite{forkel} evaluated also the
strange vector form factors and     
discussed to a great extent several different 
theoretical estimates.    

The SU(3) Skyrme model with pseudoscalar mesons~\cite{Parketal}
and with vector mesons~\cite{pw} 
estimated, respectively, $\mu_s=-0.13$, $\mu_s=-0.05$ 
and $\langle r^2\rangle^{Dirac}_s=-0.10\;\mbox{fm}^2$,
$\langle r^2\rangle^{Dirac}_s=0.05\;\mbox{fm}^2$.  Most recently, 
Leinweber obtained $\mu_s = -0.75\pm 0.30\;\mu_N$ 
which appears to be much larger than other models. 
    
In this paper, we aim at investigating the strange vector form factors
and related strange observables in the SU(3) 
chiral quark-soliton model ($\chi$QSM), often called semi-bosonized 
SU(3) Nambu-Jona-Lasinio model (NJL).
The model is based on the interaction of quarks with Goldstone bosons
and has been shown to be quite successful in reproducing static properties
of the baryons such as mass splitting~\cite{Blotzetal,Weigeletal},
axial constants~\cite{BlotzPraGoeke} and magnetic moments~\cite{Kim1}
and their form factors~\cite{Kim2,Kim3}.  
In a recent review~\cite{Christovetal}, 
one can easily see how well the model describe the baryonic observables. 
In particular, since the strange vector form factors are deeply 
related to the electromagnetic form factors~\cite{jaffe,Hammeretal}
being well described in $\chi$QSM,
it is quite interesting to study them in the same framework.  
  
The outline of the paper is as follows: Section II develops 
the general formalism for obtaining the strange vector form factors 
in the framework of $\chi$QSM.  Section III
presents the corresponding results and discuss them.
Section IV contains a summary and draws the conclusion of the present
work.
\section{General formalism}
In this section we briefly review the formalism of $\chi$QSM.
Details can be found in ref.~\cite{Christovetal}.
We start with the low-energy partition function in Euclidean space
given by the functional integral over pseudoscalar meson ($\pi^a$) and 
quark fields($\psi$):
\begin{eqnarray}
{\cal Z} &=& \int {\cal D} \psi {\cal D} \psi^\dagger {\cal D}
\pi^a \exp{\left( -\int d^4x \psi^\dagger iD \psi\right)},  
\nonumber \\
& = & \int {\cal D} \pi^a \exp{(-S_{eff}[\pi])},
\label{Eq:action}
\end{eqnarray}
where $S_{eff}$ is the effective action 
\begin{equation}
S_{eff}[\pi] \;=\;-\mbox{Sp} \mbox{ln}iD.
\end{equation}
$iD$ represents the Dirac differential operator
\begin{equation}
iD \;=\; \beta (- i \rlap{/}{\partial} + \hat{m} + MU^{\gamma5})
\end{equation}
with the pseudoscalar chiral field
\begin{equation}
U^{\gamma_5}\;=\; \exp{(i\pi^a \lambda^a \gamma_5)}\;=\;
\frac{1+\gamma_5}{2} U + \frac{1-\gamma_5}{2}U^\dagger.
\end{equation}
$\hat{m}$ is the matrix of the current quark mass given by
\begin{equation}
\hat{m} \;=\;\mbox{diag} (m_u,m_d,m_s) 
\;=\; m_0{\bf 1} \;+\; m_8 \lambda_8,
\label{Eq:mass}
\end{equation}
where $\lambda^a$ designate the usual Gell-Mann matrices normalized as 
$\mbox{tr}(\lambda^a\lambda^b) = 2 \delta^{ab}$.  
Here, we have assumed isospin symmetry ($m_u=m_d$).  
$M$ stands for the 
dynamical quark mass arising from the spontaneous chiral
symmetry breaking, which is in general momentum-dependent~\cite{dp}. 
We regard $M$ as a constant and introduce the proper-time regularization
for convenience.  The $m_0$ and $m_8$ in Eq.~(\ref{Eq:mass}) 
are defined, respectively,
by
\begin{equation}
m_0\;=\; \frac{m_u+m_d+m_s}{3},\;\;\;\;\;m_8\;=\; 
\frac{m_u+m_d-2m_s}{2\sqrt{3}}.
\end{equation}
 The operator $i D$ is expressed in Euclidean space in terms of
the Euclidean time derivative $\partial_\tau$
and the Dirac one--particle Hamiltonian $H(U^{\gamma_5})$
\begin{equation}
i D \; = \; \partial_\tau \; + \; H(U^{\gamma_5}) 
+ \beta\hat{m} - \beta\bar{m}{\bf 1} 
\label{Eq:Dirac} 
\end{equation}
with
\begin{equation}
H(U^{\gamma_5}) \; = \; \frac{\vec{\alpha}\cdot \nabla}{i} 
\;+\; \beta MU^{\gamma_5} \;+\;\beta \bar{m}{\bf 1}.
\label{Eq:hamil}
\end{equation}
$\bar{m}$ is defined by $(m_u+m_d)/2 = m_u = m_d$.
$\beta$ and $\vec{\alpha}$ are the well--known 
Dirac Hermitian matrices.  
The $U$ is assumed to have a structure corresponding to the so-called
trivial embedding of the SU(2)-hedgehog into SU(3):
\begin{equation}
U\;=\; \left ( \begin{array}{cc}
U_0 & 0 \\ 0 & 1 \end{array} \right ),
\label{Eq:imbed}
\end{equation}
with
\begin{equation}
U_0\;=\;\exp{[i\vec{n}\cdot\vec{\tau}P(r)]} .
\end{equation}
The profile function $P(r)$ is determined numerically by solving the
Euler-Lagrange equation corresponding to 
$\frac{\delta S_{eff}}{\delta P(r)}=0$.  This yields a selfconsistent
classical field $U_0$ and a set of single quark energies and corresponding
states $E_n$ and $\Psi_n$.  
Note that the $E_n$ and $\Psi_n$ do not constitute
the nucleon $|N\rangle$ yet because the collective spin and and isospin
quantum numbers are missing.  
Those are obtained by the semiclassical quantization procedure,
described below in the context of the strange form factors.

The information of the strange vector form factors 
in the nucleon is contained in the quark matrix 
elements as follows:
\begin{equation}
\langle N'(p')| J^{s}_{\mu} | N(p) \rangle \;=\;
\langle N'(p')| \bar{s} \gamma_\mu s | N(p) \rangle.
\label{Eq:mat}
\end{equation}
The strange Dirac form factors of the nucleon are defined by the
matrix elements of the $J^{s}_{\mu}$:
\begin{equation}
\langle N'(p')| J^{s}_{\mu} | N(p) \rangle\;=\;
\bar{u}_{N} (p') \left[\gamma_\mu F^{s}_1 (q^2) + i \sigma_{\mu\nu}
\frac{q^\nu}{2M_N} F^{s}_2 (q^2)\right] u_N (p),
\end{equation}
where $q^2$ is the square of the four momentum transfer $q^2=-Q^2$
with $Q^2>0$.  $M_N$ and $u_N(p)$ stand for the nucleon mass and
its spinor, respectively.
The strange quark current $J^{s}_{\mu}$ can be expressed in terms
of the baryon current and the hypercharge current:
\begin{equation}
J^{s}_{\mu}\;=\; \bar{s} \gamma_{\mu} s 
\;=\; J^{B}_{\mu} - J^{Y}_{\mu}\;=\;\bar{q}\gamma_\mu\hat{Q}_s q,
\end{equation}
where
\begin{eqnarray}
J^{B}_{\mu} & = & \frac{1}{N_c} \bar{q} \gamma_\mu q, \nonumber \\
J^{Y}_{\mu} & = & \frac{1}{\sqrt{3}} 
\bar{q} \gamma_\mu \lambda_8 q \nonumber \\
\hat{Q}_s & = &\frac{1}{N_c}
-\frac{1}{\sqrt{3}}\lambda_8,
\end{eqnarray}
where $N_c$ denotes the number of colors of the quark. 
$\hat{Q}_s=\mbox{diag}(0,0,1)$ is called {\em strangeness operator}:
We employ the non-standard sign convention
used by Jaffe~\cite{jaffe} for the strange current.
The baryon and hypercharge currents are equal to
the singlet and octet currents, respectively. 

The strange Dirac form factors $F^{s}_1$ and $F^{s}_{2}$ 
can be written in terms of the strange Sachs form factors, 
$G^{s}_E(Q^2)$ and $G^{s}_M(Q^2)$:
\begin{eqnarray}
G^{s}_E (Q^2) &=& F^{s}_1 (Q^2) - \frac{Q^2}{4M^2_{N}} F^{s}_{2}(Q^2)
\nonumber \\
G^{s}_M (Q^2) &=& F^{s}_1 (Q^2) + F^{s}_{2}(Q^2).
\end{eqnarray}
In the non--relativistic limit($Q^2 \ll M^{2}_N$), 
the Sachs-type form factors $G^{s}_{E}(Q^2)$ and $G^{s}_{M} (Q^2)$
are related to the time and space components of the strange
current, respectively:
\begin{eqnarray}
\langle N'(p') |J^{s}_{0}(0) | N(p) \rangle & = &  
G^{s}_{E} (Q^2) \nonumber \\
\langle N'(p') | J^{s}_{i}(0) | N(p) \rangle & = &  
\frac{1}{2M_N} G^{s}_{M} (Q^2) i 
\epsilon_{ijk} q^j \langle \lambda'| \sigma_k | \lambda \rangle,
\label{Eq:gm}
\end{eqnarray}
where $\sigma_k$ stand for Pauli spin matrices.  The $| \lambda\rangle$
is the corresponding spin state of the nucleon.
The matrix elements of the strange quark current can be related to
a correlator:
\begin{equation}
\langle N'(p')| \bar{s}\gamma_{\mu} s | N(p) \rangle 
\smash{\mathop{\sim}\limits_{T\rightarrow \infty}}
\langle 0 | J_{N'} (\vec{x},T/2) \bar{q} \gamma_{\mu} \hat{Q}_s q 
J^{\dagger}_{N} (\vec{y},-T/2) |0 \rangle.
\label{Eq:expect}
\end{equation}    
The nucleon current $J_N$ can be built from $N_c$ quark fields 
\begin{equation}
J_N(x)\;=\; \frac{1}{N_c !} \epsilon_{i_{1} \cdots i_{N_c}} 
\Gamma^{\alpha_1 \cdots
\alpha_{N_c}}_{JJ_3TT_3Y}\psi_{\alpha_1i_1}(x)
\cdots \psi_{\alpha_{N_c}i_{N_c}}(x).
\label{Eq:corr}
\end{equation}
$\alpha_1 \cdots\alpha_{N_c}$ denote spin--flavor indices, while
$i_1 \cdots i_{N_c}$ designate color indices.  The matrices 
$\Gamma^{\alpha_1 \cdots\alpha_{N_c}}_{JJ_3TT_3Y}$ are taken to endow 
the corresponding current with the quantum numbers $JJ_3TT_3Y$.
In our model, Eq. (\ref{Eq:expect}) is represented 
by the Euclidean functional integral with regards to quark 
and pseudo-Goldstone fields:
\begin{eqnarray}
\langle N'(p')| \bar{q}\gamma_{\mu} \hat{Q}_s q | N(p) \rangle
& = & \frac{1}{\cal Z} \lim_{T\rightarrow \infty} 
\exp{(ip_4 \frac{T}{2}
- ip'_{4} \frac{T}{2})} \nonumber \\
& \times & \int d^3 x d^3 y 
\exp{(-i \vec{p'} \cdot \vec{y} + i \vec{p} \cdot \vec{x})} 
\int {\cal D}U \int {\cal D} \psi \int {\cal D}\psi^\dagger 
\nonumber \\
& \times & \; J_{N'}(\vec{y},T/2)q^\dagger(0) 
\beta \gamma_\mu \hat{Q}_s q(0) J^{\dagger}_{N} (\vec{x}, -T/2) 
\nonumber \\ & \times &
\exp{\left[ - \int d^4 z \psi^\dagger i D \psi \right ]},
\label{Eq:stff1}
\end{eqnarray}
where ${\cal Z}$ stands for the normalization factor
which is expressed by the same functional integral but without
the quark current operator $\bar{s} \gamma_\mu s$.  
Since $\bar{m}$ is much smaller
than $m_s$, we use $\hat{m} -\bar{m}{\bf 1}\simeq diag(0,0,m_s)$ 
in the perturbation.
Eq.(\ref{Eq:stff1}) can be decomposed into valence and sea contributions:
\begin{equation}
\langle N'(p')| \bar{q}\gamma_{\mu} \hat{Q}_s q | N(p) \rangle\;=\;
\langle N'(p')| \bar{q}\gamma_{\mu} \hat{Q}_s q | N(p) \rangle_{val}\; +\; 
\langle N'(p')| \bar{q}\gamma_{\mu} \hat{Q}_s q | N(p) \rangle_{sea},
\end{equation}
where 
\begin{eqnarray}
\langle N'(p')| V_\mu(0) | N(p) \rangle_{val} & = & 
\frac{1}{\cal Z} 
\Gamma^{\beta_1 \cdots \beta_{N_c}}_{J'J'_3T'T'_3Y'}
\Gamma^{\alpha_1 \cdots \alpha_{N_c}*}_{JJ_3TT_3Y} 
\lim_{T \rightarrow \infty} \exp{(ip_4 \frac{T}{2}
- ip'_{4} \frac{T}{2})}
\nonumber \\
& \times & \int d^3x d^3y \exp{(-i\vec{p'}\cdot\vec{y}
+ i \vec{p} \cdot \vec{x})} \nonumber \\
& \times & \int {\cal D}U \exp{(-S_{eff})} 
\sum^{N_c}_{i=1} \; _{\beta_i}\langle\vec{y}, {\mbox T}/2|
\frac{1}{i D} | 0,t_z \rangle_{\gamma} 
[\beta \gamma_\mu \hat{Q}_s ]_{\gamma \gamma'} \nonumber \\ & \times &
_{\gamma'}\langle 0,t_z | \frac{1}{i D} 
| \vec{x}, -{\mbox T}/2\rangle_{\alpha_i} 
\prod^{N_c}_{j \neq i}
\; _{\beta_j} \langle \vec{y},{\mbox T}/2 |
\frac{1}{i D} |\vec{x}, -{\mbox T}/2 \rangle_{\alpha_j}
\label{Eq:val1}
\end{eqnarray}
and 
\begin{eqnarray}
\langle N'(p')| V_\mu(0) |N(p) \rangle_{sea} & = & 
-\frac{N_c}{\cal Z} 
\Gamma^{\beta_1 \cdots \beta_{N_c}}_{J'J'_3T'T'_3Y'}
\Gamma^{\alpha_1 \cdots \alpha_{N_c}*}_{JJ_3TT_3Y} 
\lim_{T \rightarrow \infty} \exp{(ip_4 \frac{T}{2}
- ip'_{4} \frac{T}{2})}
\nonumber \\
& \times & \int d^3x d^3y \exp{(-i\vec{p'}\cdot\vec{y}
+ i \vec{p} \cdot \vec{x})} \nonumber \\
& \times & \int {\cal D}U \exp{(-S_{eff})}
{\rm Tr}\ _{\gamma \lambda} 
\langle 0, t_z | \frac{1}{i D}[\beta \gamma_\mu] 
\hat{Q}_s | 0, t_z \rangle        \nonumber \\ 
& \times & \prod^{N_c}_{i=1}\; _{\beta_i} 
\langle \vec{y},{\mbox T}/2 |   \frac{1}{i D} 
| \vec{x}, -{\mbox T}/2\rangle_{\alpha_i}.
\label{Eq:sea1}
\end{eqnarray}
$S_{eff}$ is the effective chiral action expressed by
\begin{equation}
S_{eff} \;=\; -N_c {\mbox{Sp}} \mbox{ln}\left 
[ \partial_\tau \;+\; H(U^{\gamma_5})
\; + \; \beta \hat{m} \;-\;\beta\bar{m}{\bf 1}\right ].
\end{equation}

 In order to perform the collective quantization,
we have to integrate Eqs.~(\ref{Eq:val1}) and (\ref{Eq:sea1}) over 
small oscillations of the pseudo-Goldstone field around
the saddle point Eq.~(\ref{Eq:imbed}).  This will not be done
except for the zero modes.  The corresponding fluctuations of the
pion fields are not small and hence cannot be neglected.
 The zero modes are relevant to continuous symmetries
in our problem.  In particular,   
we have to take into account the translational
zero modes properly in order to evaluate form factors,  
since the soliton is not invariant under translation and its translational
invariance is restored only after integrating over the translational
zero modes.  Explicitly, the zero modes are taken into account by 
considering a slowly {\em rotating} and {\em translating} hedgehog:
\begin{equation}
\tilde{U}(\vec{x}, t)\;=\; A(t) 
U(\vec{x}-\vec{Z}(t)) A^{\dagger} (t).
\label{Eq:rot}
\end{equation}
$A(t)$ belongs to an SU(3) unitary matrix.
 The Dirac operator $i\tilde{D}$ in Eq.~(\ref{Eq:Dirac}) can be written
as
\begin{equation} 
i \tilde{D} \; = \; \left(\partial_\tau \; + \; H(U^{\gamma_5}) 
\;+\;A^{\dagger} (t) \dot{A}(t)
\;-\; i \beta \dot{\vec{Z}} \cdot \nabla
\;+\; \beta A^{\dagger} (t) (\hat{m}-\bar{m}{\bf 1}) A(t) \right).
\end{equation}
The corresponding collective action is expressed by
\begin{eqnarray}
\tilde{S}_{eff} & = & -N_c {\rm Sp}\
\mbox{ln}\left [ \partial_\tau \;+\; H(U^{\gamma_5}) 
\;+\; A^{\dagger} (t) \dot{A}(t)
\;-\;  i \beta \dot{\vec{Z}} \cdot \nabla \right .
\nonumber \\  
&  & \left .
\; + \; \beta A^{\dagger}(t) (\hat{m}-\bar{m}{\bf 1}) A(t) 
\;-\; \beta A^{\dagger}(t) s_\mu \gamma_\mu \hat{Q}_s A(t) \right ] 
\label{Eq:effact}
\end{eqnarray}
with the angular velocity 
\begin{equation}
A^{\dagger}(t)\dot{A}(t) \;=\; i\Omega_E \;=\; 
\frac{1}{2} i \Omega^{a}_{E} \lambda^a 
\end{equation}
and the velocity of the translational motion
\begin{equation}
\dot{\vec{Z}}\;=\; \frac{d}{dt} \vec{Z} .
\end{equation} 
  
Hence, Eq.~(\ref{Eq:val1}) 
and Eq.~(\ref{Eq:sea1}) can be written in terms of
the rotated Dirac operator $i\tilde{D}$ and chiral effective action
$\tilde{S}_{eff}$.     
 The functional integral over the pseudoscalar field
$U$ is replaced by the path integral which can be calculated
in terms of the eigenstates of the Hamiltonian 
corresponding to the collective action 
and these Hamiltonians can be diagonalized in an exact manner.

We take into account the rotational $1/N_c$ corrections and
$m_s$ corrections up to linear order: 
\begin{equation}
\frac{1}{i\tilde{D}}\simeq \frac{1}{\partial_\tau + H}
+\frac{1}{\partial_\tau + H}(-i\Omega_E)\frac{1}{\partial_\tau + H}
+\frac{1}{\partial_\tau + H}
(-\beta A^\dagger [\hat{m}-\bar{m}{\bf 1}]A) \frac{1}{\partial_\tau + H}.
\end{equation}
When the $m_s$ corrections are considered, SU(3) symmetry is no more
exact.  Thus, the eigenfunctions of the collective Hamiltonian are 
neither in a pure octet nor in a pure decuplet 
but in mixed states with higher representations:
\begin{equation}
| 8, N \rangle \;=\; | 8,N \rangle \;+ \; 
c_{\bar{10}} | \bar{10},N \rangle
\;+\;c_{27} | 27,N \rangle
\label{Eq:wfc}
\end{equation}
with
\begin{equation}
c_{\bar{10}} \;=\; \frac{\sqrt{5}}{15}(\sigma - r_1)I_2 m_s,
\;\; c_{27} \;=\; \frac{\sqrt{6}}{75}(3\sigma + r_1 - 4r_2)
I_2 m_s.
\label{Eq:g2}
\end{equation}
The constant $\sigma$ is related to the SU(2) $\pi N$ sigma term 
$\Sigma_{SU(2)}\;=\;3/2 (m_u + m_d) \sigma$ and $r_i$ designates
$K_i/I_i$, where $K_i$ stand for the anomalous moments of inertia
defined in~Ref.\ \cite{Blotzetal}.

Having carried out a lengthy manipulation 
(for details, see Ref.\cite{Christovetal}), 
we arrive at our final expressions for the strange vector form factors.  
The Sachs strange electric form factor $G^{s}_{E}$ is expressed as follows
 (see appendix A for detail):
\begin{eqnarray}
G^{s}_{E}(\vec{Q}^2)& = & (1 -
\langle D^{(8)}_{88}\rangle_N) {\cal B} (Q^2)
\nonumber \\
&+& \langle D^{(8)}_{8a} J_a\rangle_N 
\frac{2{\cal I}_1 (Q^2)}{\sqrt{3} I_1}    
+ \langle D^{(8)}_{8p} J_p\rangle_N 
\frac{2{\cal I}_2 (Q^2)}{\sqrt{3} I_2}    
 \nonumber \\
& + &  (1 -\langle D^{(8)}_{88}\rangle_N) m_s {\cal C} (Q^2)
\nonumber \\
& + & 
\langle D^{(8)}_{8a} D^{(8)}_{8a}\rangle_N
\frac{4 m_s}{3I_1} \left (I_1 {\cal K}_1 (\vec{Q}^2)
-{\cal I}_1 (\vec{Q}^2) K_1\right ) \nonumber \\
& + & 
\langle D^{(8)}_{8p} D^{(8)}_{8p}\rangle_N
\frac{4 m_s}{3I_2} \left (I_2 {\cal K}_2 (\vec{Q}^2)
-{\cal I}_2 (\vec{Q}^2) K_2 \right ),
\label{Eq:elecf}  
\end{eqnarray}
 $I_i$ and $K_i$ are the moments of inertia and anomalous
moments of inertia~\cite{Blotzetal}, respectively,
${\cal B}$, ${\cal I}_i$, and ${\cal K}_i$
correspond to the baryon number, moments of inertia,
and the anomalous moments of inertia
at $Q^2=0$, respectively.  
From Eq.(\ref{Eq:elecf}), we can see easily that at $Q^2=0$ the strange
electric form factor $G^{s}_{E}$ vanishes (note that 
${\cal C}(Q^2=0)=0$).  Making use of the relation 
$\sum^{8}_{a=1}D^{(8)}_{8a} J_a = -\sqrt{3}Y/2$ and
$J_8 = -N_c /(2\sqrt{3})$, we obtain
$G^{s}_{E}(Q^2=0)=B-Y=S$.  Since the net strangeness
of the nucleon is zero, $G^{s}_{E}$ at $Q^2=0$ must
vanish.  
The final expression of the Sachs strange magnetic form factor is
written (see appendix A for detail) by
\begin{eqnarray}
G^{s}_{M} (\vec{Q}^2) & = & \frac{M_N}{|\vec{Q}|} 
\left [-\frac{\langle D^{(8)}_{83}\rangle_N}{\sqrt{3}} 
\left({\cal Q}_0 (\vec{Q}^2) 
\; +\; \frac{{\cal Q}_1(\vec{Q}^2)}{I_1}
\; +\; \frac{{\cal Q}_2(\vec{Q}^2)}{I_2}\right) \right.
\nonumber \\
&+& \langle (D^{(8)}_{88}-1) J_3\rangle_N  
\frac{{\cal X}_1 (\vec{Q}^2)}{3 I_1} 
\;+\;   \langle d_{3pq}D^{(8)}_{8p}J_q \rangle_N \delta_{pq}
\frac{{\cal X}_2 (\vec{Q}^2)}{\sqrt{3} I_2} 
\nonumber \\
&-& \frac{2m_s}{\sqrt{3}} 
\langle (D^{(8)}_{88} - 1)D^{(8)}_{83}\rangle_N 
{\cal M}_0 (\vec{Q}^2) \nonumber \\
& + &  \frac{m_s}{3}  \langle D^{(8)}_{83} \rangle_N
\left(2 {\cal M}_1 (\vec{Q}^2)  \;-\; 
\frac{2}{\sqrt{3}} r_1 {\cal X}_1 (\vec{Q}^2)  \right) 
\nonumber \\
& - &  \frac{m_s}{\sqrt{3}}  
\langle D^{(8)}_{83} D^{(8)}_{88} \rangle_N
\left(2 {\cal M}_1 (\vec{Q}^2)  \;-\; 
\frac{2}{3} r_1 {\cal X}_1 (\vec{Q}^2)  \right) 
\nonumber \\
& - & \left . m_s  
\langle d_{3pq}D^{(8)}_{8p}D^{(8)}_{8q} \rangle_N\delta_{pq}
\left(2 {\cal M}_2 (\vec{Q}^2) 
\; - \;  
\frac{2}{3} r_2 {\cal X}_2 (\vec{Q}^2)  \right) \right ],
\label{Eq:magf}
\end{eqnarray}
\section{Results and Discussions}
In order to evaluate Eqs. (\ref{Eq:elecf},\ref{Eq:magf})
numerically, we follow the Kahana-Ripka discretized basis 
method~\cite{kr}.
However, note that it is of great importance to use a reasonably
large size of the box ($D\approx 10 \;\mbox{fm}$) so as to get
a numerically stable results.    
The present SU(3) $\chi$QSM (equivalent to SU(3) NJL on the chiral 
circle) contains four free parameters.  Two of them are fixed in
the meson sector by adjusting them to the pion mass, 
$m_\pi=139\ \mbox{MeV}$, the pion decay constant, 
$f_\pi=93\ \mbox{MeV}$, and the kaon mass, $m_K=496 \ \mbox{MeV}$.  
As for the fourth parameter, {\em i.e.} the constituent mass $M$
of up and down quarks, values around $M=420\ \mbox{MeV}$ have been used
because they have turned out to be the most appropriate one for the
description of nucleon observables and form factors 
(see ref.~\cite{Christovetal}).  In fact, $M=420\ \mbox{MeV}$ is the
preferred value, which is always used in this paper.
For the description of the baryon sector, we choose the method of Blotz 
{\em et al.}~\cite{Blotzetal} modified for a finite pion mass. 
The resulting strange current quark mass comes out around 
$m_s=180\ \mbox{MeV}$.  In order to illustrate the effect of the $m_s$
the calculations in the baryonic sector are performed with both
$m_s=0$ and finite $m_s$.  One should note that
a SU(3)-calculation with $m_s=0$ does not correspond to a SU(2) 
calculation, since the spaces, in which the collective
quantization are performed, are different.  

Figure 1 shows the strange electric form factor 
$G^{s}_{E} (Q^2)$, as the constituent quark mass $M$ is
varied from 370 MeV to 450 MeV without $m_s$ corrections.     
The strange electric form factor $G^{s}_{E}$ decreases as
$M$ increases.  As shown in Fig. 1, 
the $G^s_{E}(Q^2)$ without the $m_s$ corrections is 
rather insensitive to the constituent quark mass $M$.  
 In Fig. 2 the $G^s_{E}$ with the $m_s$ corrections
is drawn.  The $m_s$ corrections enhance the $G^s_{E}$
drastically, contributing to it almost $50\%$.

The Sachs and Dirac mean-square strange radii are, respectively, defined
by
\begin{equation}
\langle r^2\rangle^{Sachs}_{s} \;=\; -6\frac{d G_{E}^{s} (Q^2)}{dQ^2} 
\left.\right|_{Q^2=0}  ,\;\;\;\;
\langle r^2\rangle^{Dirac}_{s} \;=\; -6\frac{d F_{1}^{s} (Q^2)}{dQ^2} 
\left.\right|_{Q^2=0}
\end{equation}
We obtain $\langle r^2\rangle^{Sachs}_{s}=-0.25 \;\mbox{fm}^2$
and $\langle r^2\rangle^{Dirac}_{s}=-0.20 \;\mbox{fm}^2$
without the $m_s$ corrections in case of $M=420\;\mbox{MeV}$ and
$\langle r^2\rangle^{Sachs}_{s}=-0.35 \;\mbox{fm}^2$
and $\langle r^2\rangle^{Dirac}_{s}=-0.32 \;\mbox{fm}^2$
with them.  As seen from the above results, the $m_s$ corrections
increase the $\langle r^2\rangle_{s}$ considerably as in case of
the $G^s_{E}$.  However, the mechanism of the enhancement of the 
$\langle r^2\rangle_{s}$ by the $m_s$ corrections
is distinguished from that of the $G^{s}_{E}$ at finite momentum 
transfers.  
 Fig. 3 depicts the baryon ($B$) and hypercharge ($Y$) 
densities with $r^2$.  
The $m_s$ corrections elevate the baryon density in small $r$ region
sizably while they lessen its tail quite much in such a way 
that the baryon number is always 
kept to be one.  On the other hand, the hypercharge one is almost
not changed.  Hence, the increase of the baryon density
brings about the enhancement of the $G^{s}_{E}$, while the 
suppression of its tail increases the $\langle r^2\rangle_{s}$.
In fact, we obtain the baryon and hypercharge radii, respectively:
$\langle r^2\rangle_{B}=0.48\;\mbox{fm}^2$, 
$\langle r^2\rangle_{Y}=0.73\;\mbox{fm}^2$ 
without the $m_s$ corrections and 
$\langle r^2\rangle_{B}=0.35\;\mbox{fm}^2$, 
$\langle r^2\rangle_{Y}=0.70\;\mbox{fm}^2$ with
them. As the $m_s$ corrections are switched on, 
$\langle r^2\rangle_{B}$ is almost $30\%$ decreased and
yields correspondingly the enhancement of $\langle r^2\rangle_{s}$.
Fig. 4 illustrates the strange electric densities weighted with $r^2$.
As expected from the above discussion, the $m_s$ corrections amplify
the strange electric density.  

Fig. 5 draws the strange magnetic form factor 
without the $m_s$ corrections.  In contrast to the
$G^{s}_{E}$, the $G^{s}_{M}$ increases slowly with the increasing
constituent quark mass apart from the small $Q^2$ region (
below about $Q^2=0.2\;\mbox{GeV}^2$).  Fig. 6
shows the $G^{s}_{M}$ with the $m_s$ corrections.  It increases
as $M$ increases in the whole $Q^2$ region.  As we can see from the
comparison between these two figures, the $G^{s}_{M}$ is reduced
dramatically with the $m_s$ corrections being considered. 
In case of $M= 420\;\mbox{MeV}$, the $m_s$ corrections bring
it down almost by $40\%$.  The strange magnetic moment is defined
as the strange magnetic form factor at $Q^2=0$.  
 The strange magnetic moment we have obtained 
is $\mu_s = -0.44\ \mu_N$ in unit of the nuclear magneton.  
 Its absolute value is rather greater than in the other models
except for the recent calculation by Leinweber~\cite{Leinweber}:
$\mu_s=-0.75\pm 0.30\ \mu_N$.  
Fig. 7 shows the strange magnetic densities with $r^2$.

In table 1, the strange magnetic moments $\mu_s$ and
mean-square strange radii $\langle r^2\rangle_{s}$
are displayed as a function of $M$
and $m_s$.  In table 2, we have made a comparison for the $\mu_s$ and 
$\langle r^2\rangle^{Sachs}_{s}$ between different models.

We want to take the occasion to comment on Ref.~\cite{tueb}.
Though Ref.~\cite{tueb} seems to use the same model as the present
work, there are significant differences between
these two papers.  First, Weigel {\em et al.}~\cite{tueb} do not
consider rotational $1/N_c$ corrections in contrast to the present paper.
This has the immediate consequence that the magnetic moments of 
Weigel {\em et al.} are $\mu_p = 1.06\;\mu_N,\;
\mu_n=-0.69\;\mu_N$ for the nucleon
\footnote{For this comparison, the constituent quark mass
$M=450\;\mbox{MeV}$ is chosen.  In case of $M=420\;\mbox{MeV}$,
we have obtained $\mu_p=2.39\;\mu_N$ and 
$\mu_n=-1.76\;\mu_N$~\cite{Kim1}.}
whereas the present work (including those corrections) yields
$\mu_p = 2.20\;\mu_N,\;\mu_n=-1.59\;\mu_N$ with a far better comparison with
experiment ($\mu_p = 2.79\;\mu_N,\;\mu_n=-1.91\;\mu_N$).  Furthermore,
Weigel {\em et al.} regularize, besides the real part of the action,
also the imaginary one.  This meets problems in producing the anomaly
structure and is hence avoided in the approach of the present work.
In addition the calculation of Weigel {\em et al.} are not fully
self-consistent but use some scaling approximations.

\section{Summary and Conclusion}
In summary, we have calculated in the SU(3) chiral quark-soliton model
($\chi$QSM) often called the semibosonized SU(3) 
Nambu--Jona-Lasinio model, the strange electric and magnetic
form factors of the nucleon, $G^{s}_{E}(Q^2)$ and $G^{s}_{M} (Q^2)$
including the strange magnetic moment $\mu_s$, and the mean-square 
strange radius $\langle r^2\rangle_s$.  
The theory takes into account rotational $1/N_c$ corrections
and linear $m_s$ corrections.
We have obtained $\mu_s=-0.44 \ \mu_N$,
$\langle r^2\rangle^{Dirac}_s = -0.32\ \mbox{fm}^2$ and
$\langle r^2\rangle^{Sachs}_s = -0.35\ \mbox{fm}^2$.
 The results have been compared with different other models.

There are several points where the present calculations leave room for
further studies.  Apparently the dependence of the form factors
on the value of $m_s$ is quite noticeable and probably one has
to go to higher orders in perturbation theory in $m_s$.  
 The hedgehog ansatz and the embedding of SU(2) into SU(3) cause 
the asymptotic behavior of the kaon and pion fields to be similar,
which might have some influence on the form factors at low momentum
transfers.  Besides the strange vector form factors the strange
axial form factors are also of great interest.  Presently
we are performing investigations to clarify these questions.

\section*{Acknowledgment}
We would like to thank Chr.V. Christov, P.V. Pobylitsa,
M.V. Polyakov and W. Broniowski
for fruitful discussions and critical comments.
This work has partly been supported by the BMBF, the DFG
and the COSY--Project (J\" ulich).
\begin{appendix}
\section{}
In this appendix, we present all formulae appearing in 
Eqs.(\ref{Eq:elecf},\ref{Eq:magf}).    
\begin{eqnarray}
{\cal B}(\vec{Q}^2) & = & \int d^3 x \; j_0 (Qr)
\left [ \Psi^{\dagger}_{val}(x) \Psi_{val} (x) \;-\;\frac{1}{2} \sum_n
{\rm sgn} (E_n) \Psi^{\dagger}_{n}(x) 
\Psi_{n} (x) \right ], \nonumber \\ 
{\cal C}(Q^2) & = & -\frac{2 N_c}{3} \sum_{nm}\int d^3 x j_0 (Qr)
\int d^3 y \left[\frac{\Psi^\dagger (y) 
\beta \Psi_{val} (y) \Psi^{\dagger}_{val} (x)
\Psi_n (x)}{E_n - E_{val}}   \right. \nonumber \\
& & \left. \hspace{4.5cm} \;+\;\frac{1}{2} R_{\cal M} (E_n, E_m)
\Psi^{\dagger}_{n}(y) \beta \Psi_m (y) \Psi^{\dagger}_{m} (x)
\Psi_n (x)  \right], \\
{\cal I}_1 (\vec{Q}^2) & = & \frac{N_c}{6} \sum_{n, m}
\int d^3 x \;j_0 (Qr) \int d^3 y 
\left [\frac{\Psi^{\dagger}_{n} (x) \vec{\tau} \Psi_{val} (x) \cdot 
\Psi^{\dagger}_{val} (y) \vec{\tau} \Psi_{n} (y)}
{E_n - E_{val}} \right .
\nonumber \\  & & \hspace{3cm} \;+\; \left . \frac{1}{2}
\Psi^{\dagger}_{n} (x) \vec{\tau} \Psi_{m} (x) \cdot 
\Psi^{\dagger}_{m} (y) \vec{\tau} \Psi_{n} (y) 
{\cal R}_{\cal I} (E_n, E_m) \right ],
\nonumber \\
{\cal I}_2 (\vec{Q}^2) & = &\frac{N_c}{6} \sum_{n, m^{0}}
\int d^3 x \;j_0 (Qr)\int d^3 y 
\left [\frac{\Psi^{\dagger}_{m^{0}} (x) \Psi_{val} (x) 
\Psi^{\dagger}_{val} (y) \Psi_{m{^0}} (y)}
{E_{m^{0}} - E_{val}} \right .
\nonumber \\  & & \hspace{3cm} \;+\;\left . \frac{1}{2}
\Psi^{\dagger}_{n} (x) \Psi_{m^{0}} (x) 
\Psi^{\dagger}_{m^{0}} (y) \Psi_{n} (y) 
{\cal R}_{\cal I} (E_n, E_m^{0}) \right ],
\nonumber \\
{\cal K}_1 (\vec{Q}^2) & = & \frac{N_c}{6} \sum_{n, m}
\int d^3 x \;j_0 (Qr) \int d^3 y 
\left [\frac{\Psi^{\dagger}_{n} (x) \vec{\tau} \Psi_{val} (x) \cdot 
\Psi^{\dagger}_{val} (y) \beta \vec{\tau} \Psi_{n} (y)}
{E_n - E_{val}} \right .  
\nonumber \\  & & \hspace{3cm} \;+\; \left . \frac{1}{2}
\Psi^{\dagger}_{n} (x) \vec{\tau} \Psi_{m} (x) \cdot 
\Psi^{\dagger}_{m} (y) \beta \vec{\tau} \Psi_{n} (y) 
{\cal R}_{\cal M} (E_n, E_m) 
\right ],
\nonumber \\
{\cal K}_2 (\vec{Q}^2) & = & \frac{N_c}{6} \sum_{n, m^{0}}
\int d^3 x \;j_0 (Qr) \int d^3 y 
\left [\frac{\Psi^{\dagger}_{m^{0}} (x) \Psi_{val} (x) 
\Psi^{\dagger}_{val} (y) \beta \Psi_{m{^0}} (y)}
{E_{m^{0}} - E_{val}} \right .
\nonumber \\  & & \hspace{3cm} \;+\;\left . \frac{1}{2}
\Psi^{\dagger}_{n} (x) \Psi_{m^{0}} (x) 
\Psi^{\dagger}_{m^{0}} (y) \beta \Psi_{n} (y) 
{\cal R}_{\cal M} (E_n, E_m^{0}) \right ] 
\end{eqnarray}
with regularization functions
\begin{eqnarray}
{\cal R}_{I} (E_n, E_m) & = & - \frac{1}{2\sqrt{\pi}}
\int^{\infty}_{0} \frac{du}{\sqrt{u}} \phi (u;\Lambda_i) 
\left [ \frac{E_n e^{-u E^{2}_{n}} +  E_m e^{-u E^{2}_{m}}}
{E_n + E_m} \;+\; \frac{e^{-u E^{2}_{n}} - e^{-u E^{2}_{m}}}
{u(E^{2}_{n} - E^{2}_{m})} \right ],
\nonumber \\
{\cal R}_{\cal M} (E_n, E_m) & = &
\frac{1}{2}  \frac{ {\rm sgn} (E_n) 
- {\rm sgn} (E_m)}{E_n - E_m}.
\label{Eq:regul}
\end{eqnarray}

\begin{eqnarray}
{\cal Q}_0 (\vec{Q}^2) & = & 
{N_c}\int d^3 x j_1 (qr) 
\left[ \Psi ^{\dagger}_{val}(x) \gamma_{5} 
\{\hat{r} \times \vec{\sigma} \} \cdot \vec{\tau} 
\Psi_{val} (x) \right. \nonumber \\ & & 
\left . \hspace{1cm} \;-\; 
\frac{1}{2}  \sum_n {\rm sgn} (E_n) 
\Psi ^{\dagger}_{n}(x) \gamma_{5} 
\{\hat{r} \times \vec{\sigma} \} \cdot \vec{\tau} 
\Psi_{n}(x) {\cal R}(E_n)\right ],  
\nonumber \\
{\cal Q}_1 (\vec{Q}^2) & = &  \frac{iN_c}{2}\sum_{n} 
\int d^3 x j_1 (qr)
\int d^3 y \nonumber \\  & \times &
\left[{\rm sgn} (E_n)
\frac{\Psi^{\dagger}_{n} (x) \gamma_{5}
\{\hat{r} \times \vec{\sigma} \} \times \vec{\tau}   
\Psi_{val} (x) \cdot 
\Psi^{\dagger}_{val} (y) \vec{\tau} \Psi_{n} (y)}
{E_n - E_{val}} \right .
\nonumber \\  & & 
\;+\; \left . \frac{1}{2} \sum_{m}
\Psi^{\dagger}_{n} (x)\gamma_{5} \{\hat{r} \times \vec{\sigma} \} 
\times \vec{\tau}  \Psi_{m} (x) \cdot 
\Psi^{\dagger}_{m} (y) \vec{\tau} \Psi_{n} (y) 
{\cal R}_{\cal Q} (E_n, E_m) \right ],
\nonumber \\
{\cal Q}_2 (\vec{Q}^2) & = &  \frac{N_c}{2} \sum_{m^0} 
\int d^3 x j_1 (qr)
\int d^3 y \nonumber \\  & \times &
\left[ {\rm sgn} (E_{m^0})
\frac{\Psi^{\dagger}_{m^0} (x) \gamma_{5}
\{\hat{r} \times \vec{\sigma} \} \cdot \vec{\tau}   
\Psi_{val} (x) 
\Psi^{\dagger}_{val} (y) \Psi_{m^0} (y)}
{E_{m^0} - E_{val}} \right .
\nonumber \\  & + & 
\left . \sum_{n}
\Psi^{\dagger}_{n} (x)\gamma_{5} \{\hat{r} \times \vec{\sigma} \} 
\cdot \vec{\tau}  \Psi_{m^0} (x) 
\Psi^{\dagger}_{m^0} (y) \Psi_{n} (y) 
{\cal R}_{\cal Q} (E_n, E_{m^0}) \right ],
\nonumber \\
{\cal X}_1 (\vec{Q}^2) & = & N_c
\sum_{n} \int d^3 x j_1 (qr)
\int d^3 y \left[
\frac{\Psi^{\dagger}_{n} (x)\gamma_{5}
\{\hat{r} \times \vec{\sigma} \} 
\Psi_{val} (x) \cdot
\Psi^{\dagger}_{val} (y) \vec{\tau} \Psi_{n} (y)}
{E_n - E_{val}} \right .
\nonumber \\  &+ & \left. 
\frac{1}{2} \sum_{m}
\Psi^{\dagger}_{n} (x)\gamma_{5} \{\hat{r} \times \vec{\sigma} \} 
\Psi_{m} (x) \cdot 
\Psi^{\dagger}_{m} (y) \vec{\tau} \Psi_{n} (y) 
{\cal R}_{\cal M} (E_n, E_m) \right ],
\nonumber \\
{\cal X}_2 (\vec{Q}^2) & = & N_c
\sum_{m^0} \int d^3 x j_1 (qr)
\int d^3 y \left[
\frac{\Psi^{\dagger}_{m^0} (x)\gamma_{5}
\{\hat{r} \times \vec{\sigma} \} \cdot \vec{\tau} 
\Psi_{val} (x) 
\Psi^{\dagger}_{val} (y) \Psi_{m^0} (y)}
{E_{m^0} - E_{val}} \right .
\nonumber \\  & & \hspace{1cm} 
\;+\; \left . \sum_{n}
\Psi^{\dagger}_{n} (x)\gamma_{5} \{\hat{r} \times \vec{\sigma} \} 
\cdot \vec{\tau} \Psi_{m^0} (x) 
\Psi^{\dagger}_{m^0} (y) \Psi_{n} (y) 
{\cal R}_{\cal M} (E_n, E_{m^0}) \right ],
\nonumber \\
{\cal M}_0 (\vec{Q}^2) & = &  \frac{N_c}{3} \sum_{n} 
\int d^3 x j_1 (qr)
\int d^3 y \left[ 
\frac{\Psi^{\dagger}_{n} (x) \gamma_{5}
\{\hat{r} \times \vec{\sigma} \} \cdot \vec{\tau}   
\Psi_{val} (x) 
\Psi^{\dagger}_{val} (y) \beta \Psi_{n} (y)}
{E_{n} - E_{val}} \right .
\nonumber \\  & & \hspace{1cm} 
\;+\; \left . \frac{1}{2} \sum_{m}
\Psi^{\dagger}_{n} (x)\gamma_{5} \{\hat{r} \times \vec{\sigma} \} 
\cdot \vec{\tau}  \Psi_{m} (x) 
\Psi^{\dagger}_{m} (y)\beta \Psi_{n} (y) 
{\cal R}_{\beta} (E_n, E_m) \right ],
\nonumber \\
{\cal M}_1 (\vec{Q}^2) & = & \frac{N_c}{3}
\sum_{n} \int d^3 x j_1 (qr)
\int d^3 y \nonumber \\  & \times &
\left[
\frac{\Psi^{\dagger}_{n} (x)\gamma_{5}
\{\hat{r} \times \vec{\sigma} \} 
\Psi_{val} (x) \cdot
\Psi^{\dagger}_{val} (y) \beta \vec{\tau} \Psi_{n} (y)}
{E_n - E_{val}} \right .
\nonumber \\  &+ & 
\left . \frac{1}{2} \sum_{m}
\Psi^{\dagger}_{n} (x)\gamma_{5} \{\hat{r} \times \vec{\sigma} \} 
\Psi_{m} (x) \cdot 
\Psi^{\dagger}_{m} (y) \beta \vec{\tau} \Psi_{n} (y) 
{\cal R}_{\beta} (E_n, E_m) \right ],
\nonumber \\
{\cal M}_2 (\vec{Q}^2) & = &  \frac{N_c}{3} \sum_{m^0} 
\int d^3 x j_1 (qr)
\int d^3 y \nonumber \\  & \times &
\left[ 
\frac{\Psi^{\dagger}_{m^0} (x) \gamma_{5}
\{\hat{r} \times \vec{\sigma} \} \cdot \vec{\tau}   
\Psi_{val} (x) 
\Psi^{\dagger}_{val} (y)\beta \Psi_{m^0} (y)}
{E_{m^0} - E_{val}} \right .
\nonumber \\  &+ & 
\left . \sum_{n}
\Psi^{\dagger}_{n} (x)\gamma_{5} \{\hat{r} \times \vec{\sigma} \} 
\cdot \vec{\tau}  \Psi_{m^0} (x) 
\Psi^{\dagger}_{m^0} (y) \beta \Psi_{n} (y) 
{\cal R}_{\beta} (E_n, E_{m^0}) \right ]  .
\label{Eq:mdens}
\end{eqnarray}
The regularization functions for the $G^{s}_M$ are 
\begin{eqnarray}
{\cal R} (E_n) & = & \int \frac{du}{\sqrt{\pi u}} 
\phi (u;\Lambda_i) |E_n| e^{-uE^{2}_{n}},
\nonumber \\
{\cal R}_{\cal Q} (E_n, E_m) & = & \frac{1}{2\pi} c_i
\int^{1}_{0} d\alpha \frac{\alpha (E_n + E_m) - E_m}
{\sqrt{\alpha ( 1 - \alpha)}} 
\frac{\exp{\left (-[\alpha E^{2}_n + (1-\alpha)E^{2}_m]/
\Lambda^{2}_i  \right)}}{\alpha E^{2}_n + (1-\alpha)E^{2}_m},
\nonumber \\
{\cal R}_{\beta}  (E_n, E_m) & = & 
\frac{1}{2\sqrt{\pi}} \int^{\infty}_{0} 
\frac{du}{\sqrt{u}} \phi (u;\Lambda_i) 
\left[ \frac{E_n e^{-uE^{2}_{n}} - E_m e^{-uE^{2}_{m}}}
{E_n - E_m}\right].
\label{Eq:regulm}
\end{eqnarray}
The cutoff parameter 
$\phi(u;\Lambda_i)=\sum_i c_i \theta 
\left(u - \frac{1}{\Lambda^{2}_{i}} \right)$ is  
fixed by reproducing the pion decay constants and other mesonic properties
\cite{Christovetal}.
\end{appendix}
\begin{table}
\caption{The strange magnetic moments and mean-square strange radius
as varying the constituent quark mass.}
\begin{tabular}{c|c|c|c|c|c|c}
$M$ & \multicolumn{2}{c|}{$370\; \mbox{MeV}\;\;\;\;\;\;\;\;$} 
& \multicolumn{2}{c|}{$420 \;\mbox{MeV}\;\;\;\;\;\;\;\;$}
&  \multicolumn{2}{c}{$450 \;\mbox{MeV}\;\;\;\;\;\;\;\;$}  \\ \cline{1-7}
$m_s$ [MeV] & 0 & 180 & 0 & 180 & 0 & 180 \\ \hline
$\mu_s[\mu_N]$ &$-0.87$ &$-0.37$& $-0.78$ & $-0.44$ & $-0.74$ &$-0.50$  \\
$\langle r^2 \rangle^{Dirac}_{s} [\mbox{fm}^2]$ 
& $-0.22$ & $-0.47$ & $-0.19$ & $-0.32$ & $-0.16$ & $-0.27$ \\
$\langle r^2 \rangle^{Sachs}_{s} [\mbox{fm}^2]$ 
& $-0.28$ & $-0.49$ & $-0.25$ & $-0.35$ & $-0.21$ & $-0.31$
\end{tabular}
\end{table}
\begin{table}
\caption{The theoretical comparison for the strange magnetic moment 
and mean-square strange radius between different models.  
$M=420\;\mbox{MeV}$ is used for the present work.}
\begin{tabular}{cccc} 
models&$\mu_s[\mu_N]$
&$\langle r^2\rangle^{Sachs}_{s}[\mbox{fm}^2]$&references \\
\hline
Jaffe & $-0.31 \pm 0.09$ & $0.14\pm 0.07$ & \cite{jaffe} \\
Hammer {\em et al.} & $-0.24 \pm 0.03$ & $0.23\pm 0.03$ & 
\cite{Hammeretal} \\ 
Koepf {\em et al.} & $-2.6\times 10^{-2}$
&$-0.97\times10^{-2}$&\cite{Koepfetal} \\
Musolf $\&$ Burkhardt & $-(0.31\rightarrow 0.40)$
&$-(2.71\rightarrow 3.23)\times10^{-2}$& \cite{musolfa} \\
Cohen {\em et al.} &$-(0.24\rightarrow 0.32)$&$-(3.99\rightarrow 4.51)
\times 10^{-2}$ & \cite{Cohenetal} \\
Forkel {\em et al.} &\phantom{}&$1.69\times 10^{-2}$ &\cite{forkel} \\
Park $\&$ Weigel&$-0.05$&0.05&\cite{pw} \\
Park {\em et al.}& $-0.13$&$-0.11$&\cite{Parketal} \\
Leinweber & $-0.75\pm 0.30$ & &\cite{Leinweber} \\ 
Alberico {\em et al.} &$-0.14$&$0.055$ & \cite{Albericoetal} \\ 
Weigel {\em et al.}&$-0.05\rightarrow 0.25$ 
&$-0.2 \rightarrow -0.1$& \cite{tueb}\\
SU(3) $\chi$QSM &$-0.44$&$-0.35$  &Present work
\end{tabular}
\end{table}

\vfill\break
\begin{center}
{\Large {\bf Figure Captions}}
\end{center}
\noindent
{\bf Fig. 1}:
The strange electric form factor $G^{s}_{E}$ as functions of $Q^2$
without the $m_s$ corections:
The solid curve corresponds to
the constituent quark mass M=420 MeV, while
dot-dashed curve draws M=370 MeV.  The dashed curve
displays the case of M=450 MeV.  The M=420 MeV is distinguished 
since all other observables of the nucleon are then basically 
reproduced in this model. 
\vspace{0.8cm}

\noindent
{\bf Fig. 2}:
The strange electric form factor $G^{s}_{E}$ as functions of $Q^2$
with $m_s=180\;\mbox{MeV}$:
The solid curve corresponds to
the constituent quark mass M=420 MeV, while
dot-dashed curve draws M=370 MeV.  The dashed curve
displays the case of M=450 MeV.  The M=420 MeV is distinguished 
since all other observables of the nucleon are then basically 
reproduced in this model. 
\vspace{0.8cm}

\noindent
{\bf Fig. 3}:
The baryon and hypercharge densities as functions of $r$.  
The solid curve draws the baryon charge density with 
$m_s=180\;\mbox{MeV}$ while the dashed one for the same density
without the $m_s$ corrections.  The dot-dashed curve designates
the hypercharge density with $m_s=180\;\mbox{MeV}$ 
while the dot-dot-dashed one for the same density
without the $m_s$ corrections.  
The constituent quark mass $M=420\;\mbox{MeV}$ is used.

\vspace{0.8cm}
\noindent
{\bf Fig. 4}:
The strange electric density $r^2\rho^{s}_{E}$ as functions of $r$:
The solid curve draws the $r^2\rho^{s}_{E}$ with $m_s=180\;\mbox{MeV}$,
whereas the dashed one displays it without the $m_s$ corrections.
The constituent quark mass $M=420\;\mbox{MeV}$ is used.
\vspace{0.8cm}

\noindent
{\bf Fig. 5}:
The strange magnetic form factor without the $m_s$ corrections 
as a function of $Q^2$:
The solid curve corresponds to the
constituent quark mass $M = 420 \;\mbox{MeV}$, while
dashed curve draws the case of M = 450 MeV.  The dot-dashed curve
displays the case of M = 370 MeV.  
The strange quark mass is taken to be $m_s= 0$.
 The M=420 MeV is distinguished 
since all other observables of the nucleon are then basically 
reproduced in this model.
\vspace{0.8cm}

\noindent
{\bf Fig. 6}:
The strange magnetic form factor with $m_s=180\;\mbox{MeV}$
as a function of $Q^2$:
The solid curve corresponds to the
constituent quark mass $M = 420 \;\mbox{MeV}$, while
dashed curve draws the case of M = 450 MeV.  The dot-dashed curve
displays the case of M = 370 MeV.  
The strange quark mass is taken to be $m_s= 180 \;\mbox{MeV}$. 
 The M=420 MeV is distinguished 
since all other observables of the nucleon are then basically 
reproduced in this model.

\vspace{0.8cm}
\noindent
{\bf Fig. 7}:
The strange magnetic density $r^2\rho^{s}_{M}$ as functions of $r$:
The solid curve draws the $r^2\rho^{s}_{M}$ with $m_s=180\;\mbox{MeV}$,
whereas the dashed one displays it without the $m_s$ corrections.
The constituent quark mass $M=420\;\mbox{MeV}$ is used.
\vspace{0.8cm}

\vfill\break
\begin{center}
{\Large {\bf Figures}}
\end{center}
\vspace{1.6cm}
\centerline{\epsfysize=2.7in\epsffile{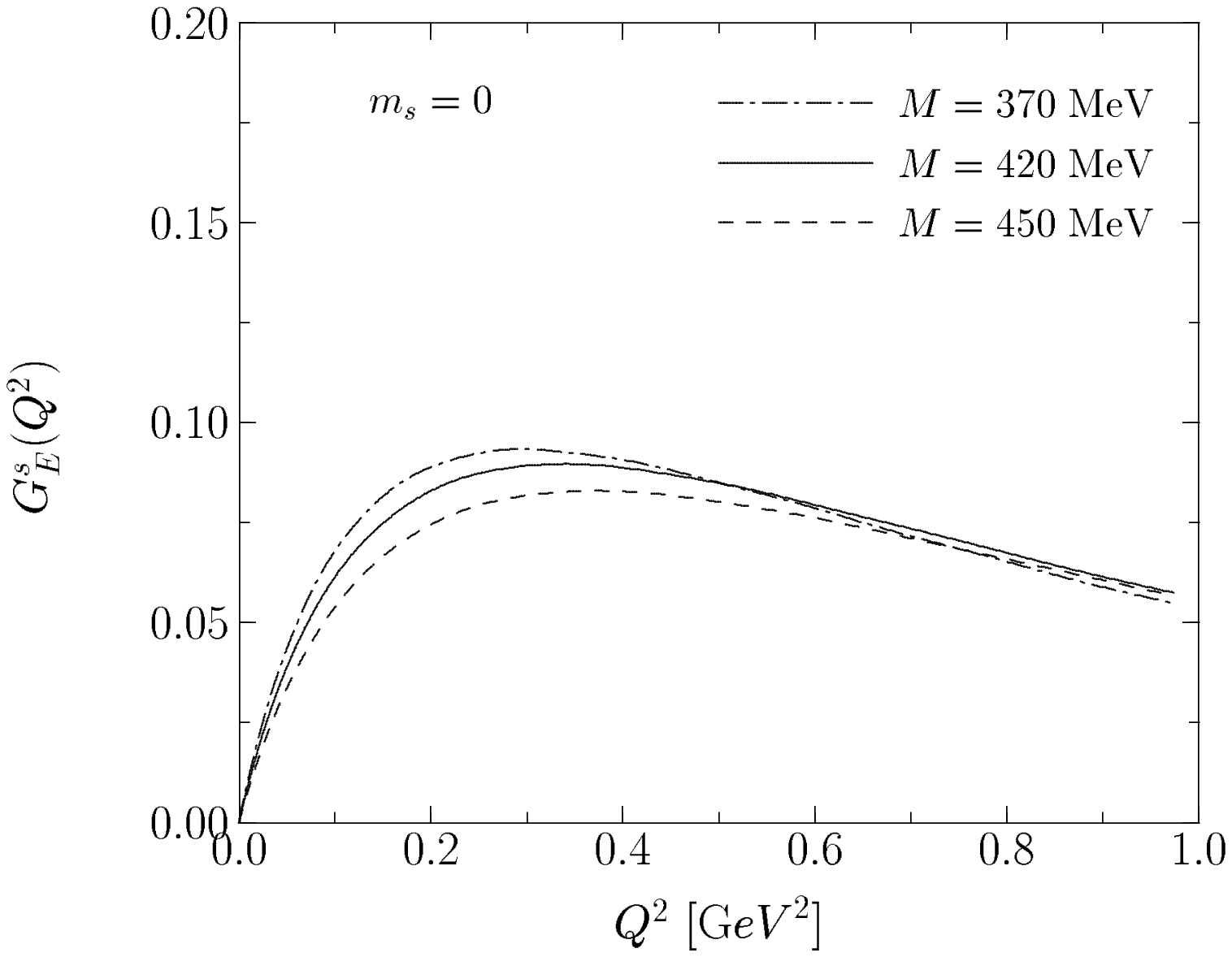}}\vskip4pt
\noindent \begin{center}	 {\bf Figure 1} 	 \end{center}   

\vspace{1.6cm}
\centerline{\epsfysize=2.7in\epsffile{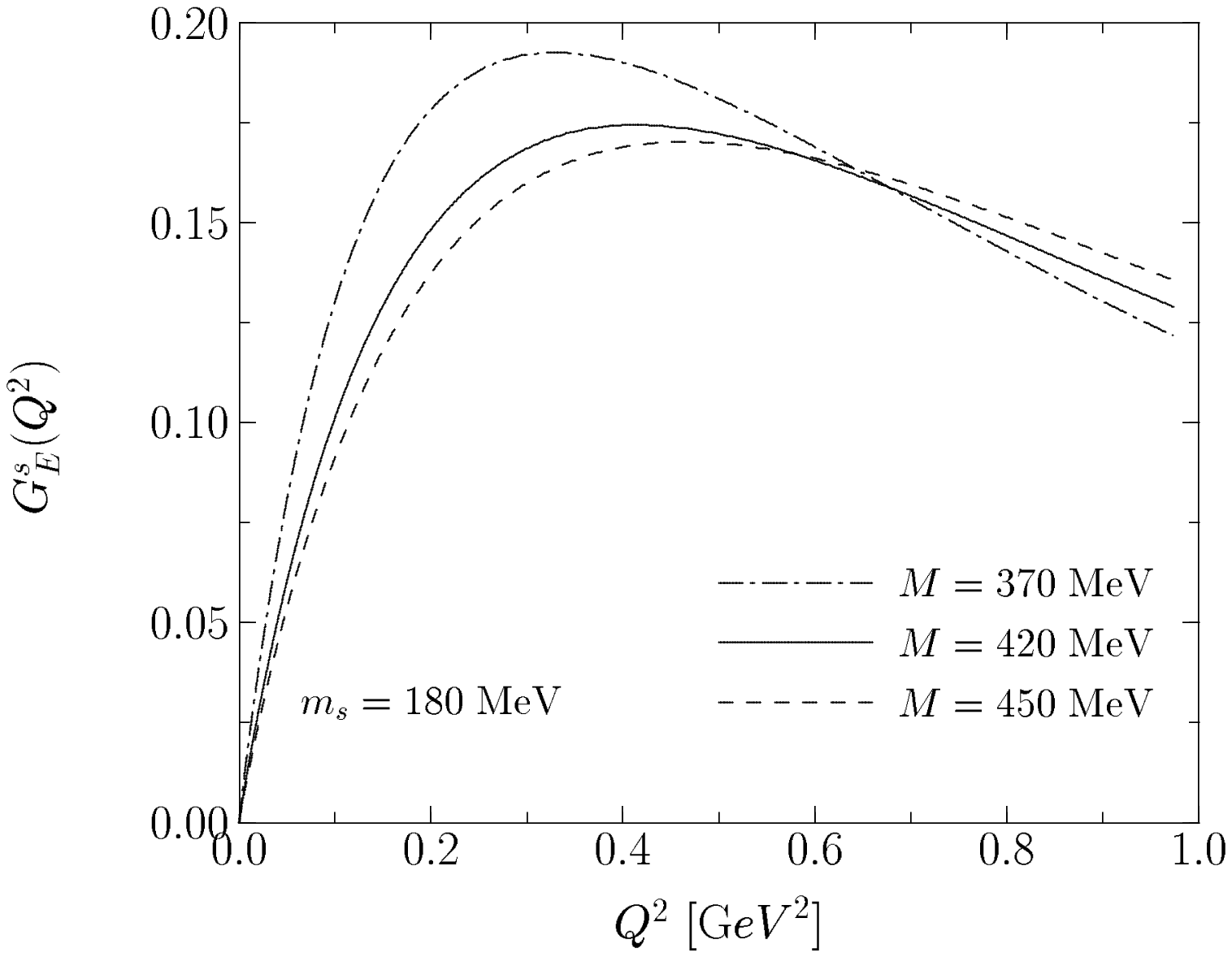}}\vskip4pt
\noindent \begin{center}	 {\bf Figure 2} 	 \end{center}   

\vspace{1.6cm}
\centerline{\epsfysize=2.7in\epsffile{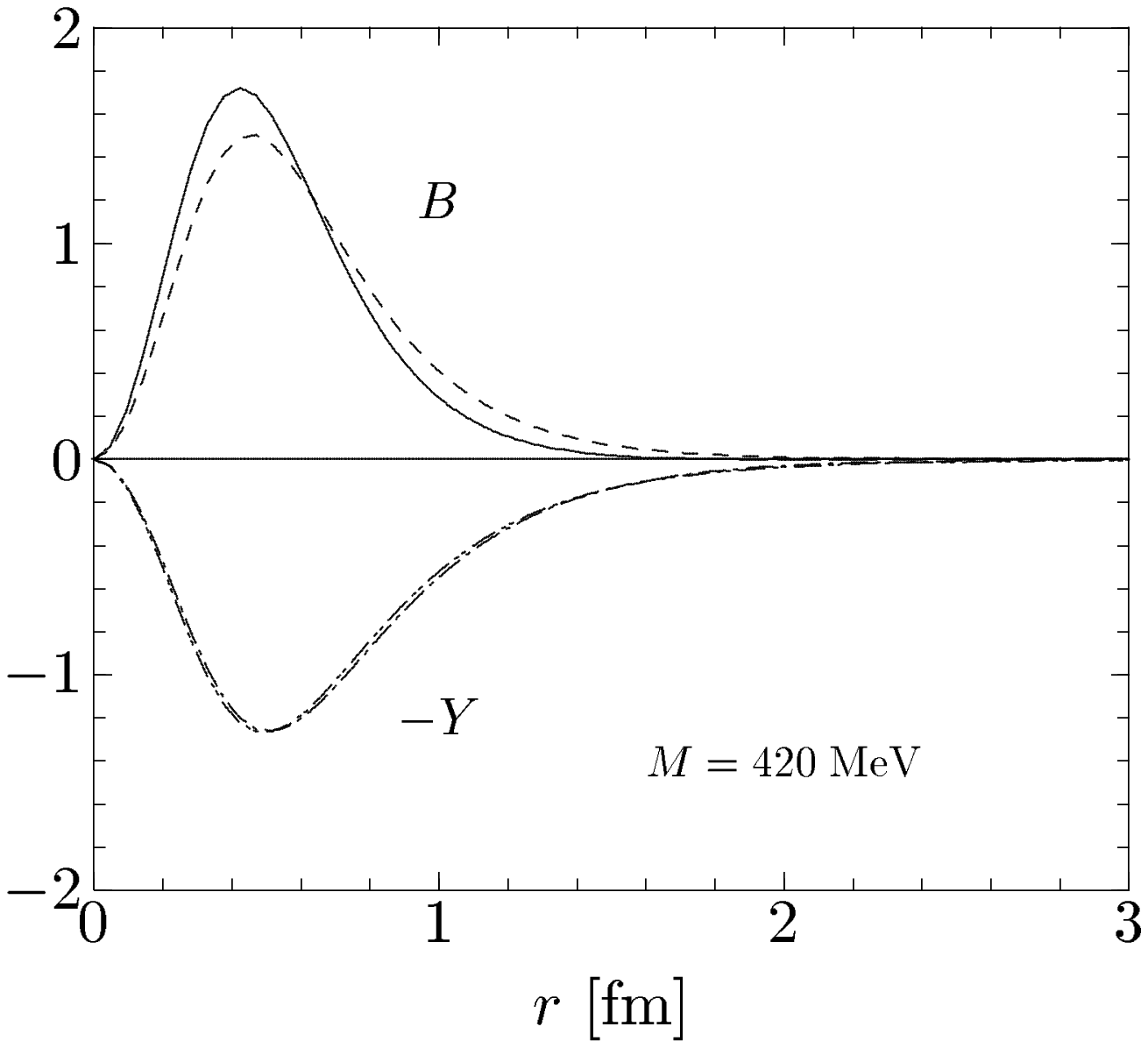}}\vskip4pt
\noindent   \begin{center}	 {\bf Figure 3} 	 \end{center}

\vspace{1.6cm}
\centerline{\epsfysize=2.7in\epsffile{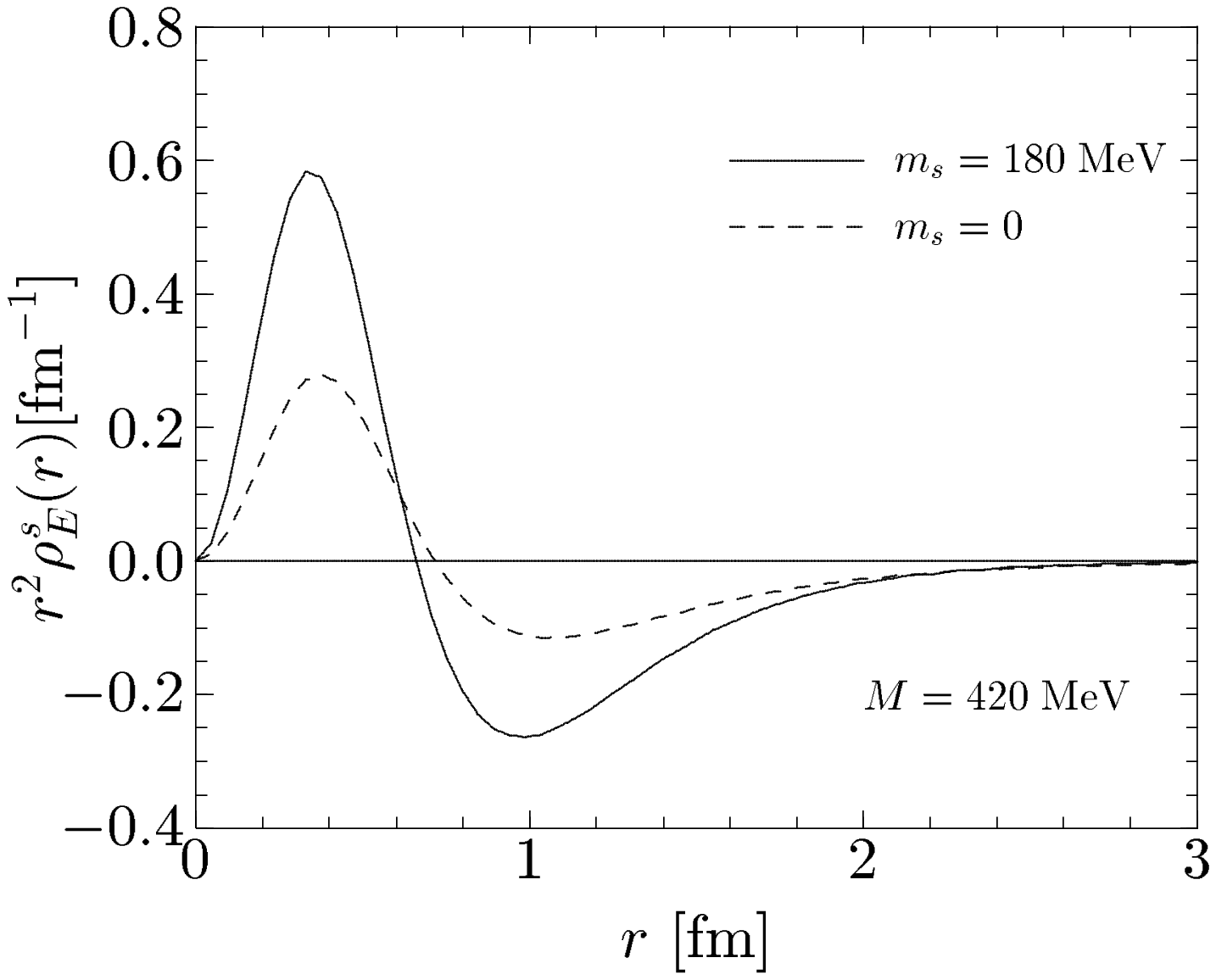}}\vskip4pt
\noindent   \begin{center}	 {\bf Figure 4} 	 \end{center} 

\vspace{1.6cm}
\centerline{\epsfysize=2.9in\epsffile{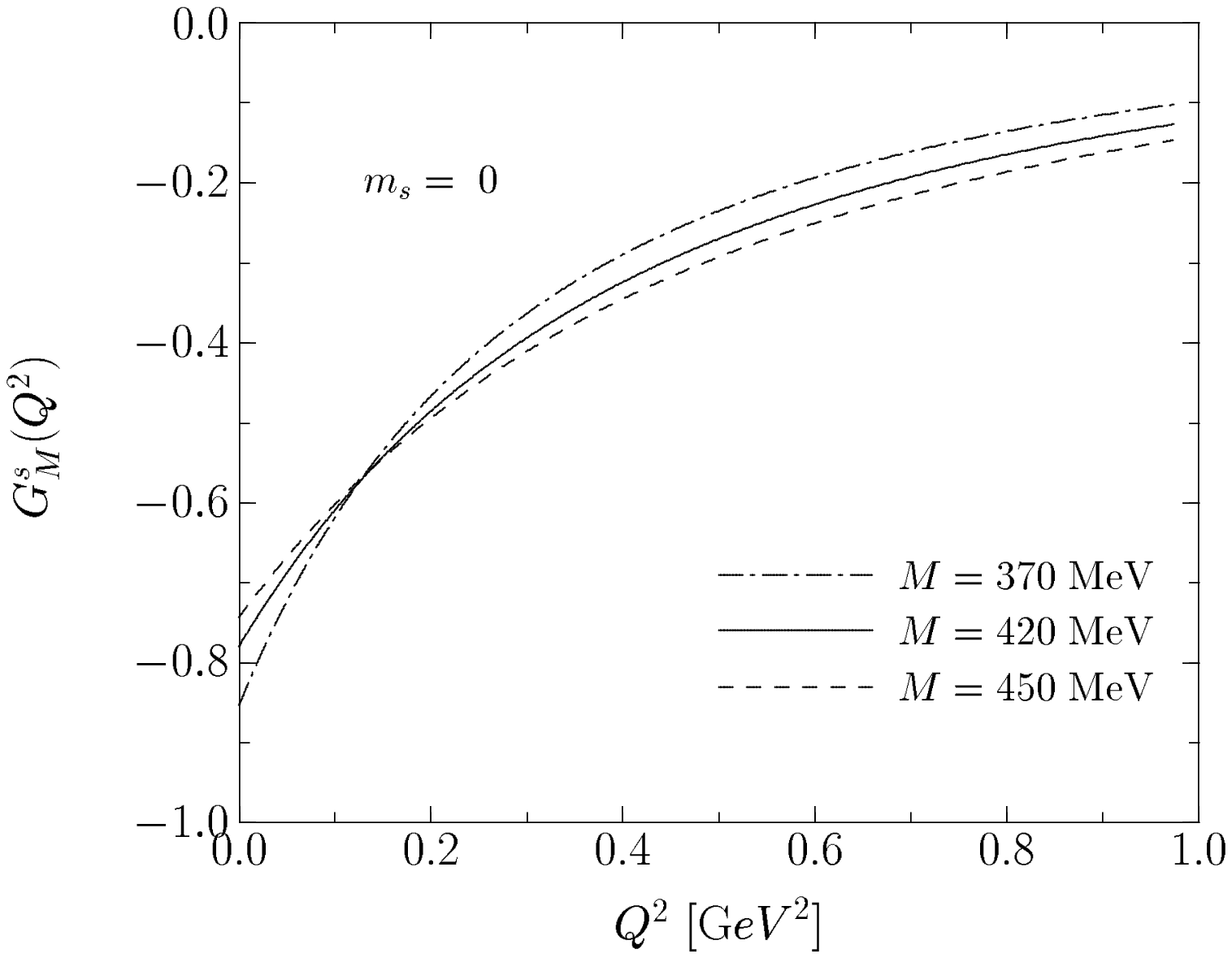}}\vskip4pt
\noindent   \begin{center}	 {\bf Figure 5} 	 \end{center} 

\vspace{1.6cm}
\centerline{\epsfysize=2.7in\epsffile{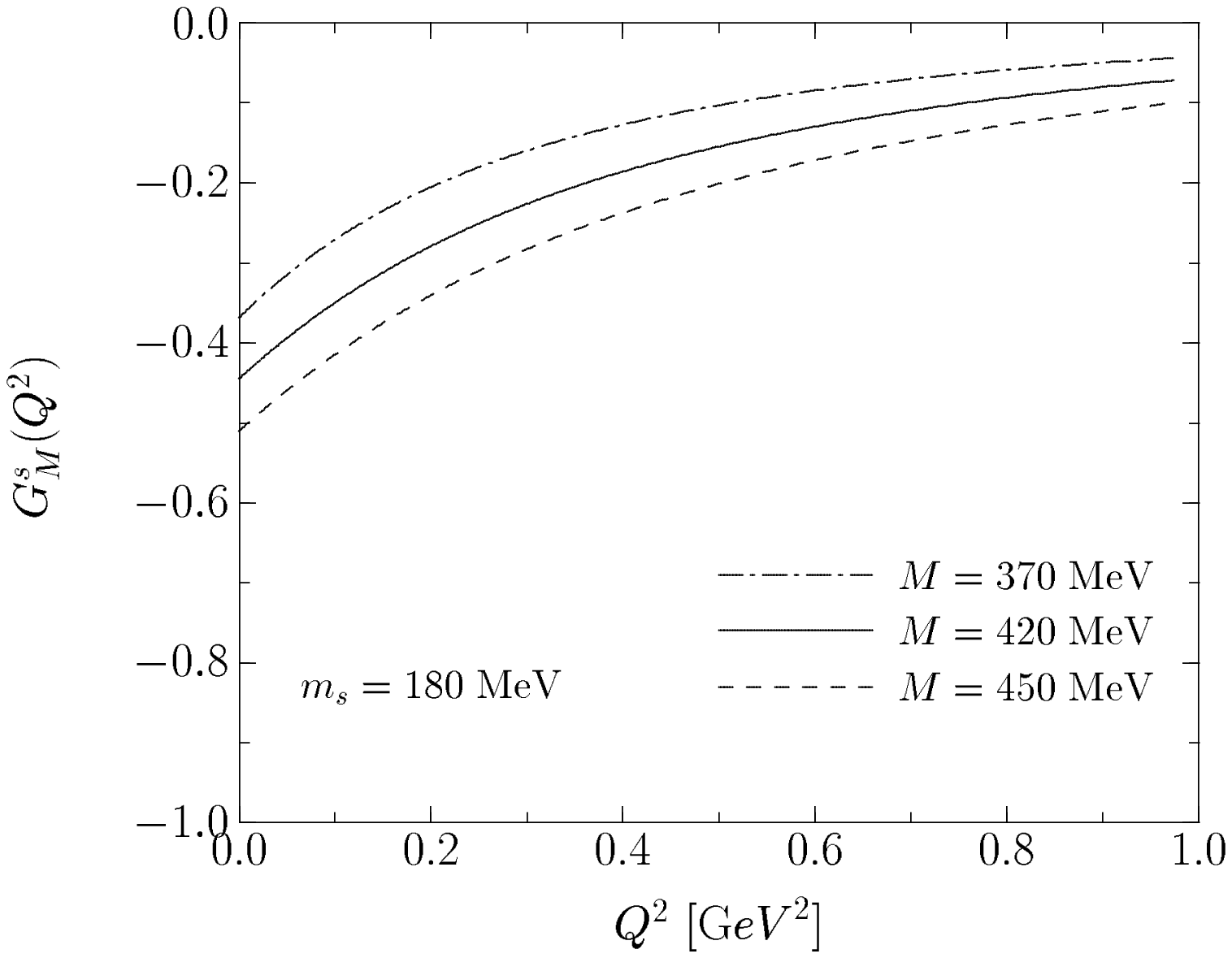}}\vskip4pt
\noindent   \begin{center}	 {\bf Figure 6} 	 \end{center} 

\vspace{1.6cm}
\centerline{\epsfysize=2.7in\epsffile{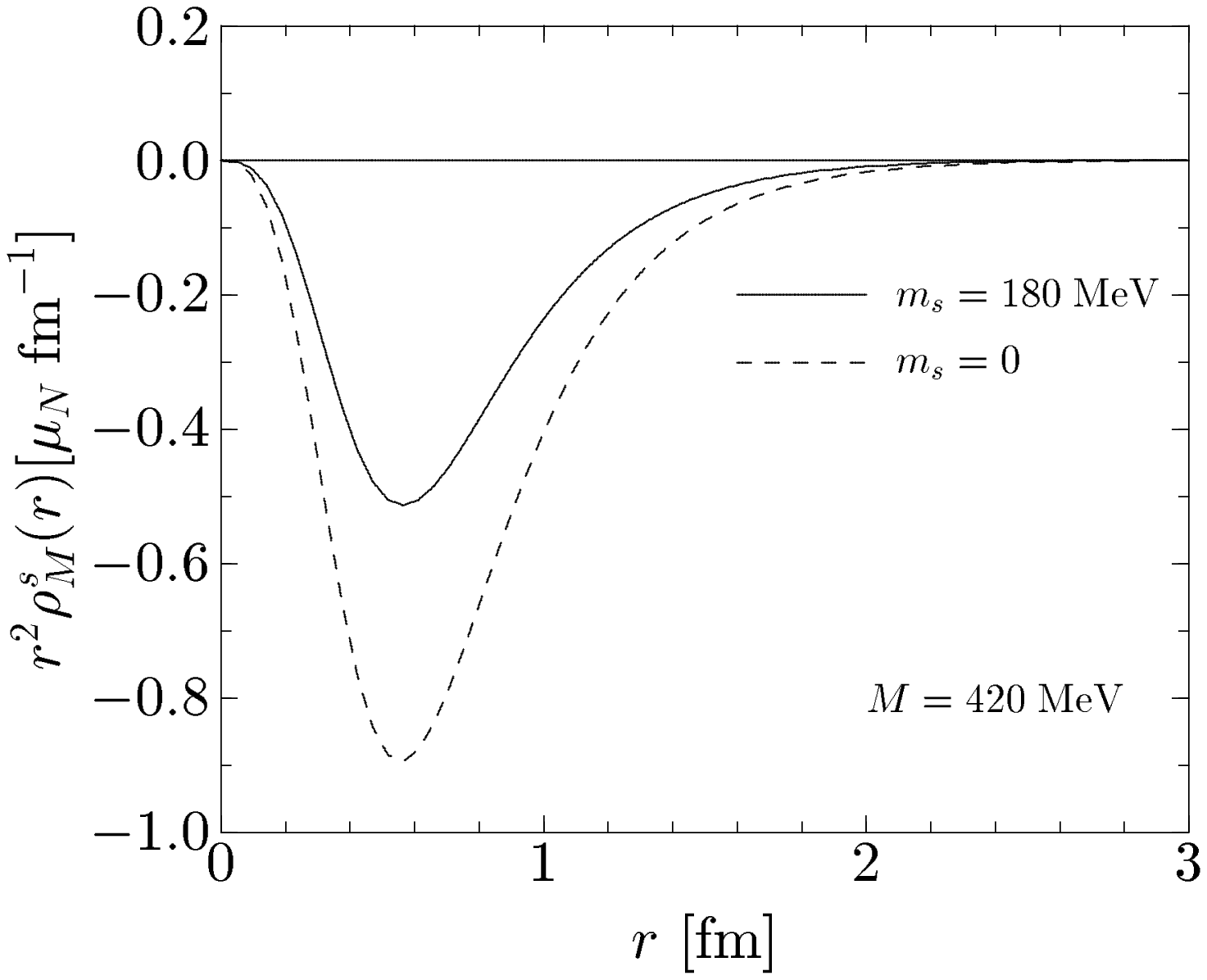}}\vskip4pt
\noindent   \begin{center}	 {\bf Figure 7} 	 \end{center}

\end{document}